\documentclass[a4paper,fleqn,usenatbib]{mnras}

\usepackage[T1]{fontenc}
\usepackage{ae,aecompl}

\usepackage{graphicx}	
\usepackage{amsmath}	
\usepackage{amssymb}	

\title[Segregation effects in DEEP2 galaxy groups]{Segregation effects in DEEP2 galaxy groups}

\author[Nascimento, Ribeiro \& Lopes]{R. S. Nascimento$^{1}$\thanks{E-mail: rnascimento@astro.ufrj.br}, A. L. B. Ribeiro$^{2}$ and P. A. A. Lopes$^{1}$\\
$^{1}$Observat\'orio do Valongo, Universidade Federal do Rio de Janeiro, RJ 20080-090, Brazil\\
$^{2}$Laborat\'orio de Astrof\'isica Te\'orica e Observacional, Universidade Estadual de Santa Cruz, Ilh\'eus 454650-000, Brazil}

\date{Accepted 2016 September 12. Received 2016 September 12; in original form 2016 April 04}

\pubyear{2015}

\begin{document}
\label{firstpage}
\pagerange{\pageref{firstpage}--\pageref{lastpage}}
\maketitle

\begin{abstract}
We investigate segregation phenomena in galaxy groups in the range of $0.2<z<1$. We study a sample of groups selected from the 4th Data Release of the DEEP2  galaxy redshift survey.  We used only groups with at least 8 members within a radius of 4$\;$Mpc. Outliers were removed with the shifting gapper techinque and, then, the virial  properties were estimated for each group. The sample was divided into two stacked systems: low($z\leq 0.6$) and high ($z>0.6$) redshift groups.  Assuming that the color index ${\rm (U-B)_0}$ can be used as a proxy for the galaxy type, we found that the fraction of blue (star-forming) objects is higher in the high-{\it z} sample, with blue objects being dominant at $M_{B} > -19.5$ for both samples, and red objects being dominant 
at $M_{B} < -19.5$ only for the low-{\it z} sample. Also, the radial variation of the red fraction indicates that there are more
red objects with $R < R_{200}$ in the low-{\it z} sample than in the high-{\it z} sample. Our analysis indicates statistical evidence of kinematic segregation, at the 99\% c.l., for the low-{\it z} sample: redder and brighter galaxies present lower velocity dispersions than bluer and fainter ones. We also find a weaker evidence for spatial segregation between red and blue objects, at the 70\% c.l.
The analysis of the high-{\it z} sample reveals a different result: red and blue galaxies have velocity dispersion distributions not statistically distinct, although redder objects are more concentrated than the bluer ones at the 95\% c.l.  From the comparison of blue/red and bright/faint fractions, and considering the approximate lookback timescale between the two samples ($\sim$3 Gyr), our results are consistent with a scenario where bright red galaxies had time to reach energy equipartition, while faint blue/red galaxies in the outskirts infall to the inner parts of the groups, thus reducing spatial segregation from $z\sim 0.8$ to $z\sim 0.4$.
\end{abstract}




\section{Introduction}

A central issue concerning galaxy formation and evolution refers to environmental factors. It is well-established that the average properties of galaxies such as their mass, colours, morphologies, and gas content depend upon the environment where they reside. Galaxies in clusters tend to be more massive and have lower star formation rates (SFRs) than isolated field galaxies which are, in general, actively star forming \citep{dr1,o1,bl,co,co1,ka,Ta, l2, l3, r2}. It is also well-known that galaxy properties depend strongly on galaxy mass \citep[e.g.][]{Pog}, and that galaxy mass and environment are correlated, since denser environments tend to be inhabited by more massive galaxies \citep[e.g.][]{HO,BG}.

Several physical processes are thought to be relevant in regulating star formation in dense environments by driving cold gas away from galaxies and by heating it up. Some of these mechanisms are more effective in dense regions like rich clusters, whereas in groups of galaxies other mechanisms play the most important role. For example, galaxy interactions as mergers and harassment are favoured in group environment because of the low relative velocities between galaxies \citep{z1}, while in high density environments galaxies can be strongly affected by mechanisms such as ram pressure stripping and strangulation due to the high temperature and pressure of the intra-cluster medium \citep[e.g.][]{vdb,Pre}. Coupled with these processes, massive galaxies tend to reduce their velocities through the energy equipartition by dynamical friction with less massive galaxies \citep{ch,Cape}.

A possible consequence of these environmental factors are the so-called segregation effects, that is, correlations between galaxy properties and/or radial trends of those properties as a function of the group/cluster centre. The presence of the segregation effect in galaxy clusters and groups has been studied by several authors. \citet{b2} studying luminosity and morphological segregation in an ensemble of 59 rich nearby clusters, observed in the ESO Nearby Cluster Survey, found that luminosity segregation is evident only for elliptical galaxies brighter than $M_R$ = -22.0$\pm$0.1, and not located in substructures. \citet{gi2} analysed morphology and luminosity segregation of galaxies in loose groups identified in the Nearby Optical Galaxy catalogue. They  concluded that spatial segregation is stronger than kinematical segregation and that luminosity is independent of morphological segregation. They argued that segregation phenomena are mainly connected with the initial conditions at the time of galaxy formation and that the mechanisms which influence galaxy luminosity and morphology should act in a similar way in groups and in clusters. \citet{la} examined a sample selected from the 2dF Galaxy Redshift Survey to analyse the segregation effect in galaxy groups. They found that passively star forming galaxies show a statistically narrower velocity distribution than that of galaxies with a substantial star forming activity. They also found that the sample of red galaxies, with  colour index B$-$R $>$ 1, have a larger fraction of small velocities ($v/\sigma<$1) compared with the blue galaxies. \citet{go1} selected a sample of 335 clusters from the Sloan Digital Sky Survey (SDSS) and found that bright cluster galaxies ($M_z < -23$) have significantly smaller velocity dispersion than fainter galaxies. They also pointed out that the results remain the same when the sample is splitted in star forming late type and passive late type galaxies, with the former having a larger velocity dispersion in comparison with the last. 
\citet{r1}, using a sample of 57 groups selected from the 2df Percolation-Inferred Galaxy Group catalogue, found that galaxies brighter than $M_R$ = -21.5 show a decrease in normalized velocity dispersion, $\sigma_u$, while for the fainter ones the velocity dispersion is approximately constant. Interestingly, the result remains for groups considered dynamically non-evolved, but with a steeper correlation between $\sigma_u$ and $M_R$. \citet{vdb}, using the SDSS group catalogue of \citet{yan}, suggest that satellite galaxies become redder and more concentrated than central galaxies once they fall into a bigger halo. However, they do not find  indication that the magnitude of the transformation depends on environment. Also using SDSS clusters, \citet{Von} find no evidence for mass segregation in four redshift bins at $z< 0.1$. A similar result is found by \citet{Vulc} using mass-limited samples at $0.3 \leq z \leq 0.8$ from the IMACS Cluster Building Survey and the ESO Distant Cluster Survey. Recently, \citet{Rob} show that failure to find mass segregation is due to a mass completeness cut at intermediate to high stellar mass, or to take only high-mass haloes. \citet{Rob} also show that mass segregation is enhanced with the inclusion of low-mass galaxies, and decreases with increasing halo mass.

Currently, few studies are available regarding segregation phenomena at intermediate and high redshifts. For instance, \citet{Pre} find evidence for mass segregation in zCOSMOS groups at both $0.2 \leq z \leq 0.45$ and $0.45 < z \leq 0.8$. By splitting up their sample into poor and rich groups at $0.2 \leq z \leq 0.45$, they find evidence for mass segregation in rich groups but not in poor groups. Also, \citet{Bal} find evidence for mass segregation in the Group Environment Evolution Collaboration 2 (GEEC2) for groups at $0.8 < z < 1$, using a stellar mass-limited sample with $M_{star}>10^{10.3}~{\rm M}_\odot$. In a recent paper, \citet{Bst} find evidence for velocity segregation in a collection of 41 galaxy clusters at $0.4 \leq z \leq 1.5$.

In the present work, we probe velocity and spatial segregation in low-mass galaxy groups, that is, the possibility of more luminous and redder galaxies being more central and move more slowly than fainter and bluer ones. Our aim is to compare these segregation phenomena for well selected samples defined in two redshift intervals, at $z\sim 0.4$ and $z\sim 0.8$. The paper is organized as follows: in Section 2 we present a description of the data used, i.e, the DEEP2 survey and group catalogue, and the method used to define the group virial properties; in Section 3 we present the main results of velocity segregation in luminosity and morphological type; in Section 4 we discuss some possible systematics; and finally in Section 5 we discuss our results. Throughout this work we assume a $\Lambda$CDM cosmology with the cosmological parameters $\Omega_M=0.3$, $\Omega_\Lambda=0.7$ and $h=0.7$. 

\section[]{data and methodology }

\subsection{DEEP2 Sample}

The DEEP2 Galaxy Redshift Survey \citep{n1} is considered the largest spectroscopic survey of homogeneously selected galaxies at $z \sim$ 1. The survey covers a total area of 2.8 deg$^2$ distributed across four fields observed up to limiting magnitude R$_{AB}$ = 24.1. Each field was chosen to lie in zones of low Galactic extinction based on the dust maps of \citet{s1}. The DEEP2 fields probe a volume of 5$\times$ 10$^6h^{-1}$Mpc$^3$ over the primary DEEP2 redshift range 0.75 $< z <$ 1.4.

The photometric catalogue for DEEP2 is derived from Canada-France-Hawaii Telescope (CFHT) images taken with the 12k$\times$8k mosaic camera \citep{c1} in B, R and I bands. DEEP2 spectroscopic observations were carried out using the 1200-line diffraction grating on DEIMOS multi-object spectrograph \citep{f1} on Keck II telescope. The spectral resolution  of R $\sim$ 6000 yielded a velocity accuracy of $\sim$ 30~km~s$^{-1}$. The typical exposure time is 1 hr per mask. The total number of spectra obtained is 52,989, and the total number of objects with secure redshift is 38,348 (DEEP2 redshift quality flag 3 or 4 which correspond to 95\% and 99\% confidence in the redshift identification, respectively).

Objects are pre-selected in DEEP2 fields 2-4 using broad-band CFHT 12k BRI photometry to remove foreground galaxies below $z \sim$ 0.7. In the DEEP2 field 1 or Extended Groth Strip \citep[EGS,][]{d1}, however, there is no rejected low-{\it z} galaxies, both to test the selection methods and to take advantage of the wide multiwavelength coverage data in that field.

K-corrections, absolute M$_{B}$ magnitudes, and rest-frame (U-B) colours have been derived as described in \citet{wi}. Absolute magnitudes presented in this paper are in the AB system and are $M_B$ - 5log {\it h} with $h = 0.7$.

\begin{figure*}  
\includegraphics[width = 180mm]{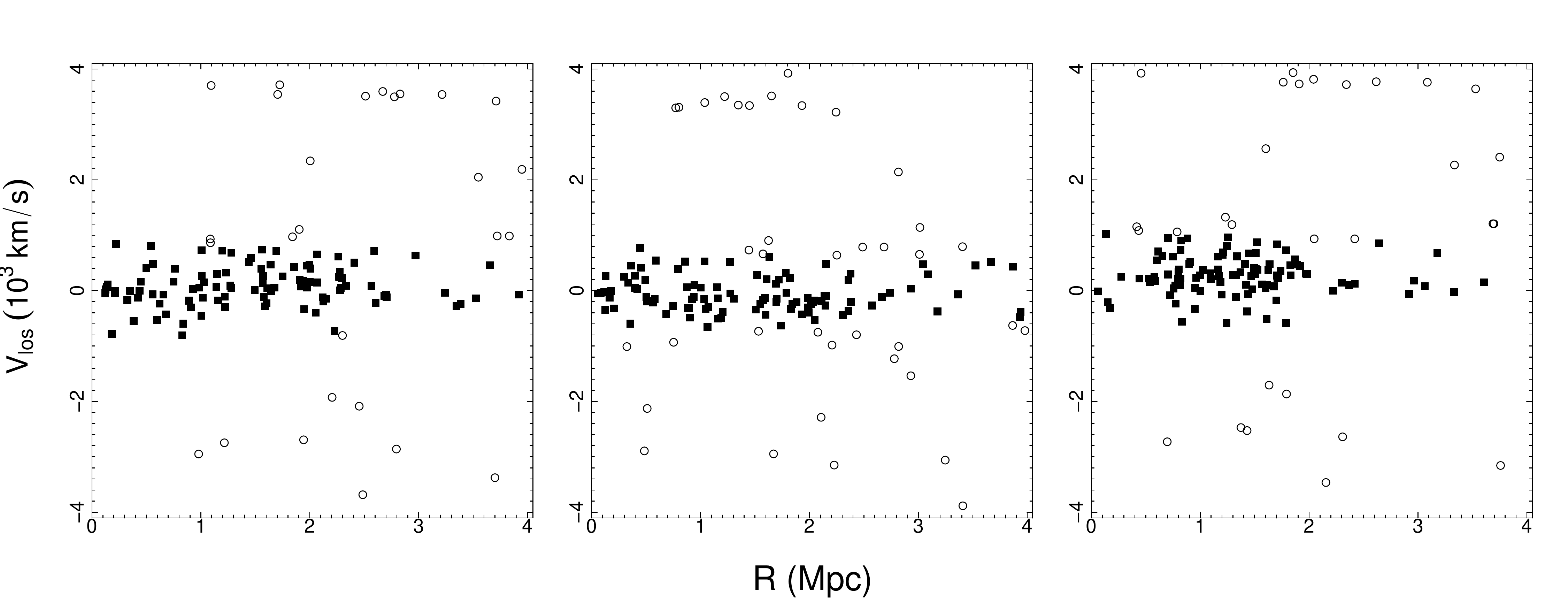}
\caption{Phase-space diagrams of 3 massive galaxy groups shown as examples. The velocity and radial offset are with respect to the group centre. We apply a shifting gapper procedure for the selection of group members (filled black squares) and exclusion of interlopers (open circles).}
\label{fig1}
\end{figure*}

\subsection{DEEP2 Group Catalogue and Virial Analysis}

This section gives a brief description of the DEEP2 group sample and for more details the reader is referred to \citet{g1}. Groups were identified using the Voronoi-Delaunay Method \citep{m1}. The algorithm yielded 1165 groups with two or more members with accurate redshifts in the EGS over the range 0 $< z <$ 1.5 and 1295 groups at $z > $ 0.6 in the rest of DEEP2. In additional to the coordinates and central redshift, the group catalogue provides estimates of the total number of galaxies in the group and its velocity dispersion. However, we only consider the positional and redshift information, re-deriving the member list and group properties (velocity dispersion, radius and mass). 

\begin{figure*}
\includegraphics[width = 130mm]{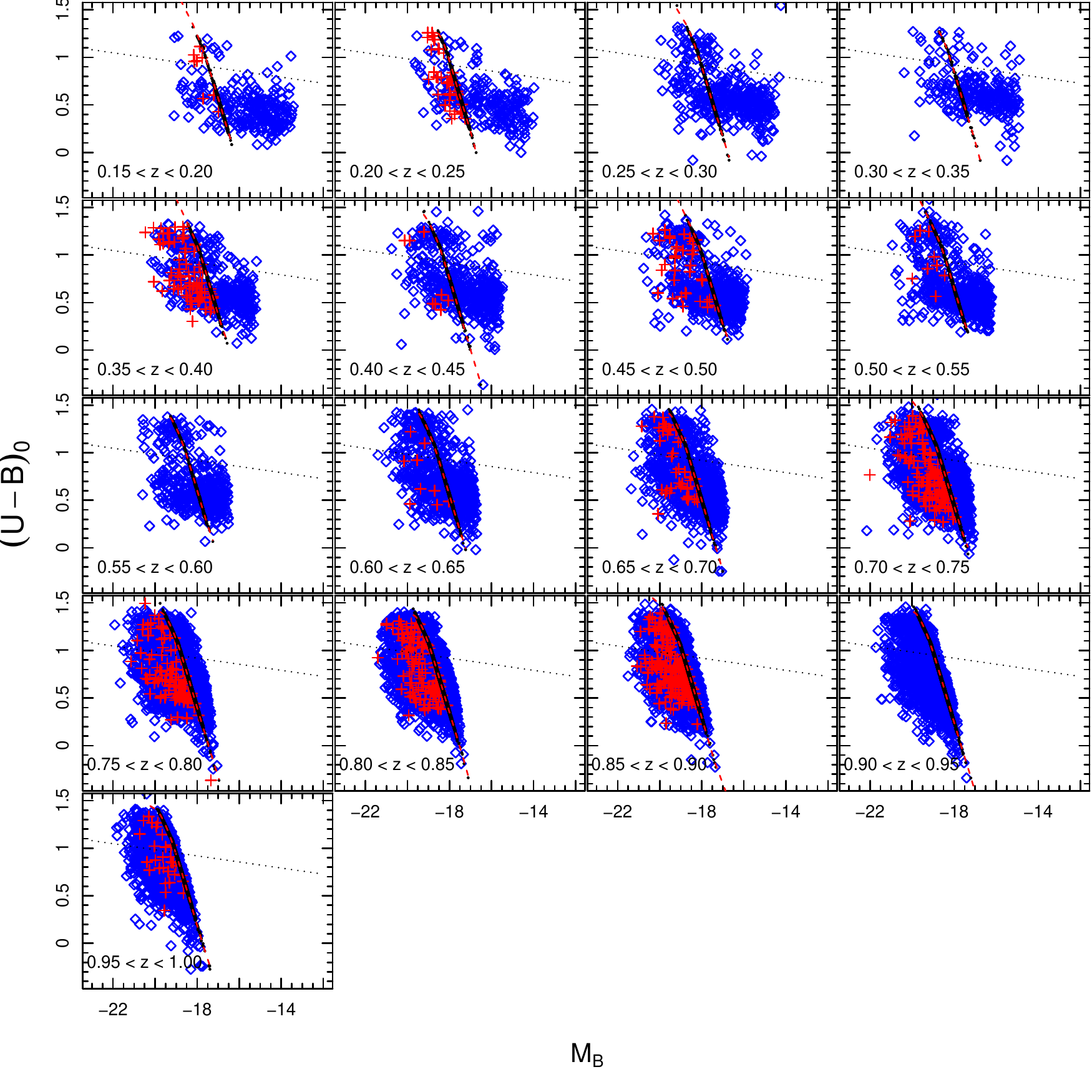}
\caption{Rest-frame colour-magnitude diagram for DEEP2 si\-mi\-lar to obtained by \citet{g2} but considering only galaxy groups with at least eight member galaxies. Each panel correspond to bins of width $\delta$z = 0.05. The red crosses and blue diamonds represent groups and field galaxies, respectively. The dashed lines show the selection cut as illustrated in equation \ref{eq1} while the dotted lines indicate the separation between red and blue galaxy population as given in equation \ref{eq2}.}
\label{fig2}
\end{figure*}

To select group members and exclude interlopers we adopted the ``shifting gapper'' technique \citep{f96, a1, l1}, using all galaxies with redshift quality 3 or 4 of the DEEP2 Data Release 4 (DR4). Around each DEEP2 group we initially considered galaxies within a maximum radius of 4 Mpc and velocity offset $|cz - cz_{group}|$~$\le$~4000~km~s$^{-1}$, where c is the speed of light, {\it z} and $z_{group}$ are galaxy and group redshifts, respectively. This large maximum radius is important to probe the effect of secondary infall on to groups.

The ``shifting gapper'' technique applies the gap technique \citep{k1, l07} in radial bins from the cluster centre. The advantage of this method is that it makes no assumption about the dynamical state of the group. For more details on the procedure we adopted see \citet{l1}. After removing interlopers, we kept the 221 groups with at least 8 member galaxies selected. Such low multiplicity allow us to explore galaxy groups in the low mass regime. Figure \ref{fig1} illustrates the procedure for three groups of our sample. In each panel, the filled black squares represent the group members and the open circles represent the rejected interlopers.

Next, we estimate the line-of-sight velocity dispersion, $\sigma_p$, for all group members. Then, we obtain an estimate of the projected virial radius (R$_{PV}$) and a first estimate of the virial mass is derived from equation 5 of \citet{gi1}. A first estimate of R$_{200}$, and a Navarro, Frenk \& White (1997; NFW) profile are assumed when applying the surface pressure correction. After that we obtain a refined estimate of R$_{200}$ considering the virial mass density. We assume again a NFW profile to obtain estimates of M$_{500}$ and M$_{200}$, and then  R$_{500}$, R$_{200}$. This procedure is analogous to \citet{bv} and \citet{l1}. The results of the virial analysis for the ten richest groups are listed, as an example, in Table \ref{tab1}. The columns represent: group name; coordinates (Right Ascension and Declination); mean redshift; velocity dispersion ($\sigma_p$); number of galaxies used to compute the velocity dispersion (N$_{\sigma}$); characteristic radii and masses (R$_{500}$, M$_{500}$, R$_{200}$, M$_{200}$). In general our groups represent low masses systems with estimates between 5 $\times 10^{12}M_\odot \leq M_{200} \leq$ 1.63 $\times 10^{14}M_\odot$.

\begin{table*}
 \centering
 \begin{minipage}{140mm}
\caption{Velocity dispersion, characteristic radii and masses of 10 of the 221 DEEP2 groups. The full table is available in eletronic form.}
\label{tab1}
\begin{tabular}{c c  c c c c c c c c}
\hline
ID &    RA    &    DEC   &  z & $\sigma_p$ & $N_{\sigma}$ & $R_{500}$ &  $M_{500}$ 		   &  $R_{200}$		&  $M_{200}$ \\
     &(J2000) & (J2000)  &    & kms$^{-1}$ &               & (Mpc)     & ${\rm (10^{14})M_\odot}$  & (Mpc)		& ${\rm (10^{14})M_\odot}$ \\

\hline\hline
\vspace{0.1cm}
1   & 215.0325 & 53.1012 & 0.2009 & 300.87{\tiny$_{-19.26}^{+27.64}$} & 104 & 0.69{\tiny$_{-0.03}^{+0.04}$} & 1.16{\tiny$_{-0.15}^{+0.21}$} & 0.95{\tiny$_{-0.04}^{+0.06}$} & 1.18{\tiny$_{-0.15}^{+0.22}$}\\
\vspace{0.1cm}
2   & 215.3142 & 53.1008 & 0.2014 & 278.44{\tiny$_{-15.24}^{+22.52}$} & 100 & 0.66{\tiny$_{-0.02}^{+0.03}$} & 1.00{\tiny$_{-0.11}^{+0.16}$} & 0.90{\tiny$_{-0.03}^{+0.05}$} & 1.03{\tiny$_{-0.11}^{+0.17}$}\\
\vspace{0.1cm}
5   & 215.1649 & 53.1322 & 0.2010 & 237.76{\tiny$_{-17.01}^{+22.77}$} & 89  & 0.58{\tiny$_{-0.03}^{+0.04}$} & 0.69{\tiny$_{-0.10}^{+0.13}$} & 0.80{\tiny$_{-0.04}^{+0.05}$} & 0.70{\tiny$_{-0.10}^{+0.13}$}\\\vspace{0.1cm}
7   & 215.2346 & 53.1516 & 0.2017 & 204.24{\tiny$_{-16.25}^{+22.31}$} & 74  & 0.52{\tiny$_{-0.03}^{+0.04}$} & 0.50{\tiny$_{-0.09}^{+0.11}$} & 0.72{\tiny$_{-0.04}^{+0.05}$} & 0.51{\tiny$_{-0.08}^{+0.11}$}\\\vspace{0.1cm}
33  & 214.9645 & 53.0123 & 0.7444 & 292.30{\tiny$_{-20.57}^{+27.68}$} & 71  & 0.60{\tiny$_{-0.03}^{+0.04}$} & 1.44{\tiny$_{-0.20}^{+0.27}$} & 0.83{\tiny$_{-0.04}^{+0.05}$} & 1.47{\tiny$_{-0.21}^{+0.28}$}\\\vspace{0.1cm}
41  & 215.3367 & 53.0569 & 0.2008 & 222.21{\tiny$_{-15.87}^{+20.03}$} & 84  & 0.55{\tiny$_{-0.03}^{+0.03}$} & 0.59{\tiny$_{-0.08}^{+0.11}$} & 0.76{\tiny$_{-0.04}^{+0.04}$} & 0.60{\tiny$_{-0.09}^{+0.11}$}\\\vspace{0.1cm}
52  & 214.3444 & 52.5873 & 0.2367 & 258.41{\tiny$_{-21.36}^{+29.95}$} & 72  & 0.72{\tiny$_{-0.04}^{+0.05}$} & 1.36{\tiny$_{-0.23}^{+0.32}$} & 0.99{\tiny$_{-0.05}^{+0.08}$} & 1.40{\tiny$_{-0.23}^{+0.32}$}\\\vspace{0.1cm}
62  & 215.1013 & 53.0645 & 0.2000 & 292.23{\tiny$_{-20.33}^{+28.64}$} & 102 & 0.67{\tiny$_{-0.03}^{+0.04}$} & 1.06{\tiny$_{-0.15}^{+0.21}$} & 0.92{\tiny$_{-0.04}^{+0.06}$} & 1.09{\tiny$_{-0.15}^{+0.21}$}\\\vspace{0.1cm}
78  & 215.1344 & 53.0142 & 0.2025 & 267.70{\tiny$_{-18.94}^{+26.69}$} & 89  & 0.65{\tiny$_{-0.03}^{+0.04}$} & 0.96{\tiny$_{-0.14}^{+0.19}$} & 0.89{\tiny$_{-0.04}^{+0.06}$} & 0.99{\tiny$_{-0.14}^{+0.20}$}\\\vspace{0.1cm}
210 & 215.0737 & 52.9612 & 0.7452 & 236.67{\tiny$_{-17.75}^{+25.39}$} & 65  & 0.53{\tiny$_{-0.03}^{+0.04}$} & 1.00{\tiny$_{-0.15}^{+0.22}$} & 0.73{\tiny$_{-0.04}^{+0.05}$} & 1.02{\tiny$_{-0.15}^{+0.22}$}\\
\hline
\end{tabular}
\end{minipage}
\end{table*}

\subsection{Galaxy Sample selection}

To define an uniform sample of galaxies in the DEEP2 redshift interval, we follow the procedure described in \citet{g2}. According to this work, in galaxy evolution studies it is possible to produce volume-limited catalogues with a colour-dependent, absolute magnitude cut by defining a region of rest-frame colour-magnitude space that is uniformly sampled by the survey at all redshifts of interest. Such a selection cut is illustrated in Fig. \ref{fig2} and is given by the equation

$$
M_{cut} - 5\log h =  Q(z - z_{lim})~~~~~~~~~~~~~~~~~~~~~~~~~~~~~~~~~~~~~~
$$
\begin{equation}
~~~~~~~~~~~~~~~~~~~~~~~~~~~~~~~~~~~ + \min\{[a(U-B)~+~b], [c(U-B)~+~d]\}, ~~~~~~~
\label{eq1} 
\end{equation}

\noindent where $z_{lim}$ is the limiting redshift beyond which the selected sample becomes incomplete; a, b, c and d are constants that depend on $z_{lim}$ and are determined by inspection of the colour-magnitude diagram; and Q is a constant that allows for linear redshift evolution of the characteristic galaxy absolute magnitude $M_B^*$. For the parameter Q, we adopt the \citet{f2} value of Q~=~-1.37, determined from a study of the B-band galaxy luminosity function in the COMBO-17 \citep{w1}. Adopting this approach and using $z_{lim}$ = 1 and consequently (a, b, c, d) = (-1.34, -18.55, -2.08, -17.77), we constructed a volume-limited  sample for each color containing 835 galaxies in the range of $0.2 \le z \le 1$ and distributed over 105 galaxy groups. 

The result of the selection cut is illustrated in Fig. \ref{fig2} which shows rest-frame colour-magnitude diagrams for DEEP2 galaxies split into redshift bins of $\Delta z$ = 0.05. The red crosses and blue diamonds represent group galaxies and field galaxies, respectively. The dashed lines show the selection cut as illustrated in equation \ref{eq1} while the dotted lines indicate the separation between red and blue galaxy populations described mathematically by

\begin{equation}
(U-B)_0 = -0.032(M_B + 21.62) + 1.035
\label{eq2} 
\end{equation}

\noindent This equation was derived from the \citet{v1} colour-magnitude relation for red galaxies in distant clusters and converted to the cosmological model used in this work. 

From now on all analyses will be made considering only galaxies whose absolute magnitude is below the completeness cut, i.e. M$_B \le M_{cut}$. It is noteworthy that, since the DEEP2 groups contain only a few members, the group properties such as velocity dispersion, characteristic radius and mass were obtained considering the full member galaxies defined after interloper removal. We choose to do this, to achieve the best statistical reliability in determining the group properties. 

\begin{figure}
\includegraphics[width = 80mm]{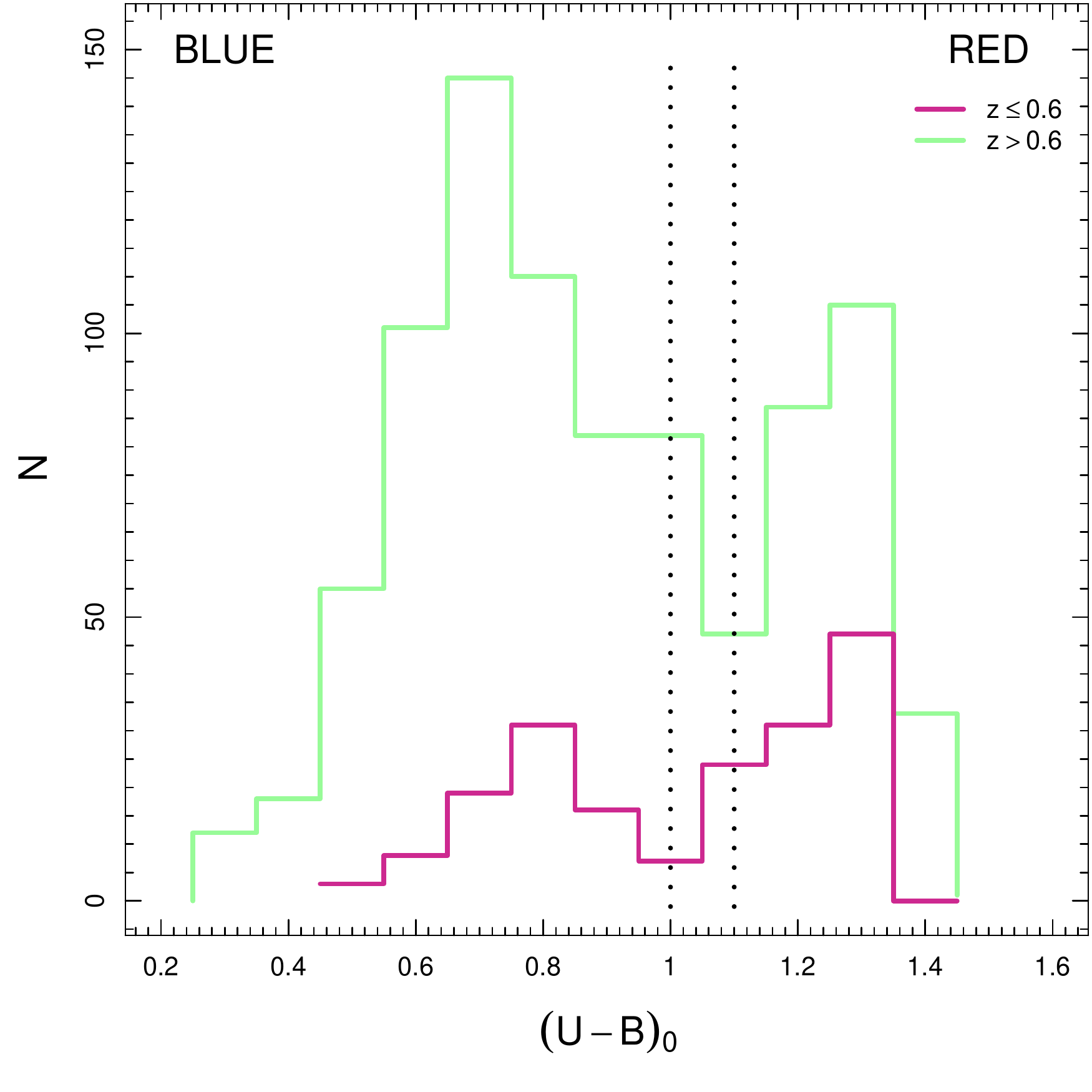}
 \caption{Histograms of $(U-B)_0$ for low-redshift (violet) and high-redshift sample (green). The vertical dotted lines indicate the separation between blue and red objects. These limits are $(U-B)_0=1.0$ and $(U-B)_0=1.1$ for low and high redshift sample, respectively.}
\label{fig3}
\end{figure}

\section{Segregation analysis}

\subsection{The composite samples}

An appropriate way to explore galaxy properties in multiple galaxy systems is to combine them in stacked samples \citep{b1,r1}. In these composite groups, the distances of galaxies to the group centres are normalized by $R_{200}$ and their velocities are referred to the group median velocities and scaled by the group velocity dispersion. The normalized velocity dispersion of the combined system, $\sigma_u$, is related to the dimensionless quantity $u_i$, defined by the equation,

\begin{equation}
u_i = \frac{v_i - \langle v \rangle_j}{\sigma_j}
\label{eq3} 
\end{equation}

\noindent where $i$ and $j$ are the galaxy and  group indices, respectively.

In order to probe the presence of luminosity se\-gre\-ga\-tion with respect to galaxy velocities, we computed $\sigma_u$ in bins of absolute magnitudes. Data allocation in bins was optimised to have approximately the same number of galaxies in each bin, so the variability of data within the bin is not size dependent. Finally, error-bars were obtained from a bootstrap technique with 1,000 resamplings. We considered galaxies with $\sigma_u <1$ (or $\sigma_u >1$) as low (or high) velocity dispersion galaxies \citep[e.g.][]{la,go1}.

The whole analysis refers to data allocated in two redshift intervals since  galaxy properties changes as we move from low to high {\it z}. An important difference between the two subsamples concerns the color range of galaxies in high and low redshifts. Galaxies in groups at high-{\it z} are, in general, predominantly bluer than those at low-{\it z}. In Fig. \ref{fig3}, we  see the $(U-B)_0$ histograms for  the two samples. The vertical dashed lines indicate the separation between blue and red objects. These limits (defined as the minimum count between the two color peaks) are $(U-B)_0=1.0$ and $(U-B)_0=1.1$ for the low and high redshift sample, respectively. This figure clearly shows the dominance of blue galaxies at high-{\it z}. These constant color cuts were tested and approximately agree with the $M_B$ dependent color cut given in equation \ref{eq2}. They are also consistent with the color cut used by \citet{Pre} for samples within a similar redshift range. In the sequence of this work, we use these color separators to characterize galaxy types.

\begin{figure}
\includegraphics[width = 80mm]{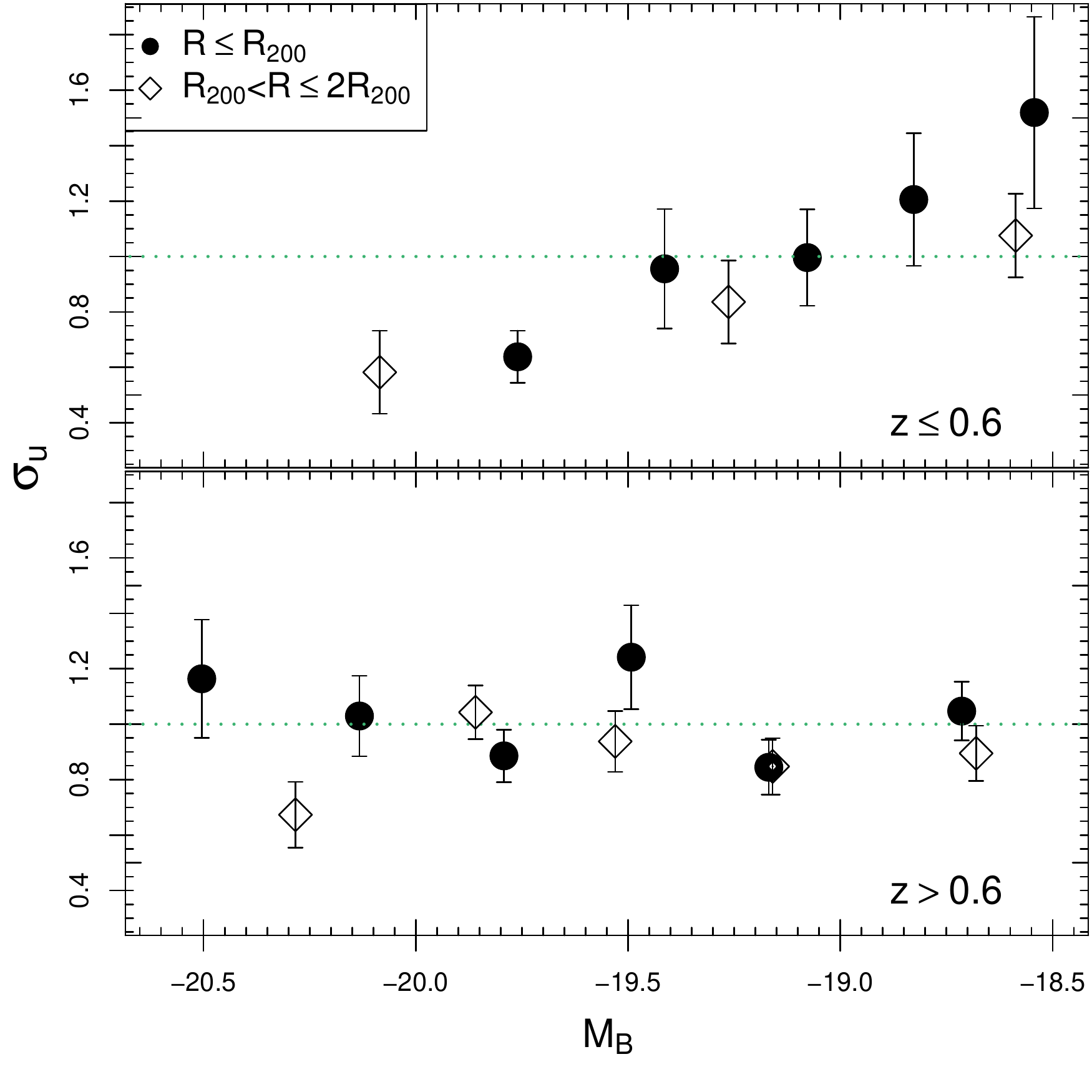}
 \caption{Normalized velocity dispersion as a function of the absolute magnitude in the B band for all galaxies in the composite group within R$_{200}$ (filled black dots) and at 1 $<$ R$_{200}$ $\leq$ 2 (empty diamond) divided in two bins of redshift. The top panel correspond to the sample at low redshift while the bottom panel represent the high redshift sample. The green lines indicate the separation between low and high velocity dispersion.}
\label{fig4}
\end{figure}

\subsection{Searching for segregation}

Using Eq. \ref{eq3} we computed the normalized velocity dispersion, $\sigma_u$, for the stacked groups in the two redshift bins. The result can be seen in Fig.~\ref{fig4} which shows $\sigma_u$ of the composite group as a function of absolute magnitude in the B band. Filled black dots and empty diamonds represent galaxies residing within $R \leq R_{200}$ and between $R_{200} < R \leq 2R_{200}$, respectively. The top panel correspond to the sample at low redshift while the bottom panel represent the high redshift sample. We also indicate in this plot the separation between low and high velocity dispersion through green lines. This threshold was adopted from \citet{la}. Looking at the low-{\it z} panel, we can see that for both $R\leq R_{200}$ and $R_{200} < R \leq 2R_{200}$ data there is a pronounced trend between $\sigma_u$  and M$_B$. Testing for association between these properties with the Pearson's correlation coefficient $\rho$ \citep[e.g][]{ed1} we find strong correlations, with coefficients $\rho=0.97$ within $R \leq R_{200}$, and $\rho=0.98$ within $R_{200} < R \le 2R_{200}$ at the 99\% confidence level. The normalized velocity dispersion for objects brighter than $M_B\approx -19.0$ is relatively smaller than the velocity dispersion of fainter objects by a factor of $\sim$0.5. This indicates that the brightest galaxies are moving more slowly than the faintest group members. On the other hand, regarding to the high redshift panel, no trend is observed between $\sigma_u$ and $M_B$ (p=0.65 within $R\leq R_{200}$ and p=0.68 within $R_{200} < R \leq 2R_{200}$ for the Pearson's correlation test). A direct interpretation of this result is just considering that the velocities of the brightest galaxies had already been reduced through dynamical interactions at $z \leq 0.6$, while that effect has not happened yet at $z > 0.6$.

\begin{figure}
\includegraphics[width=86mm,height=94mm]{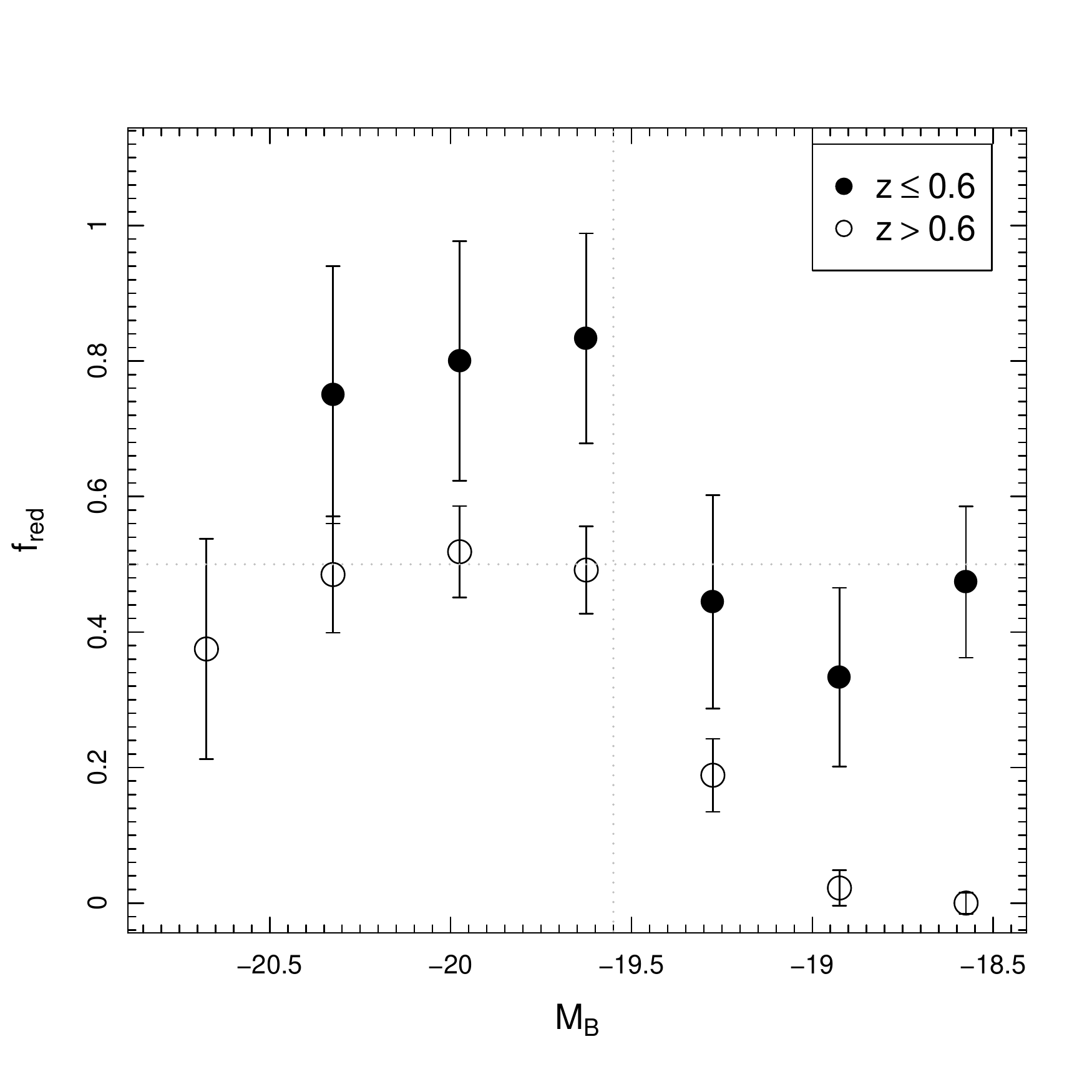}
 \caption{Fraction of red galaxies as a function of $M_B$,  taking galaxies up to
$2R_{200}$. Filled (open) circles depict galaxies at $z\leq 0.6$ ($z > 0.6$). Error bars are the standard error of the sample proportion in the binomial standard deviation. The vertical and horizontal line represent, respectively, the separation between the bright and faint red galaxy populations and  when $f_{red}$ is higher than 50\%.}
\label{fig5}
\end{figure} 

The observed trend could also be related to the distribution of galaxy types in each sample. As shown by \citet{SL}, \citet{st}, \citet{AD}, and other authors, disk galaxies in clusters have higher velocity dispersion than spheroidal galaxies.  A similar result is reached for emission line galaxies in ENACS clusters \citep{b3}, in 2dFGRS \citep{la}, and in SDSS clusters \citep{go1}. In the latter, galaxies are also classified by color and it is verified that the velocity dispersion is larger for the blue galaxies. All this indicates that the distribution of galaxy types over the absolute magnitude interval can provide a better understanding of the segregation effect observed in the upper panel of Fig.~\ref{fig4}.

\begin{figure}
\includegraphics[width=86mm,height=94mm]{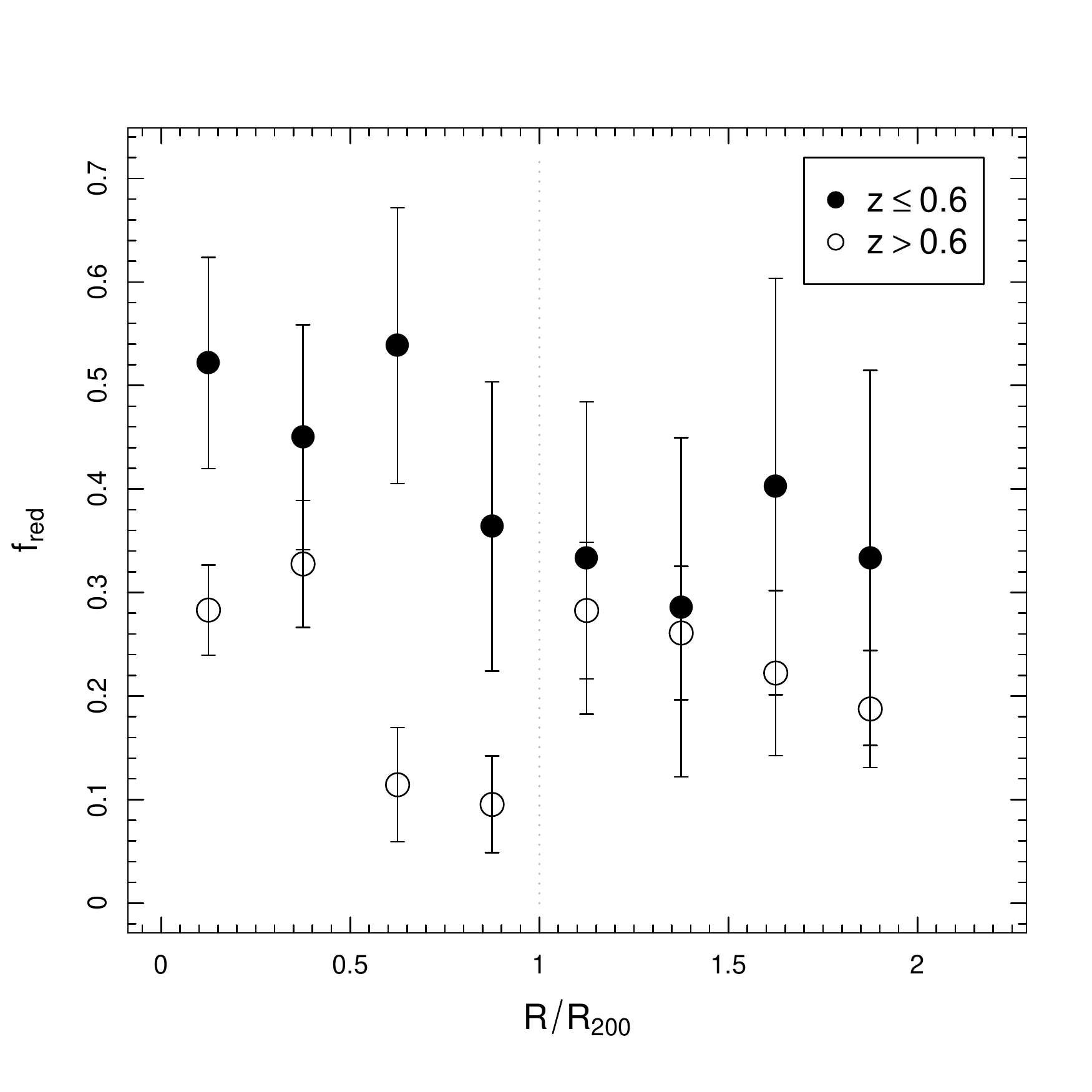}
 \caption{Fraction of red galaxies as a function of $R/R_{200}$,  taking galaxies up to $2R_{200}$ with $M_B \leq -18.5$. Filled (open) circles depict galaxies at $z\leq 0.6$ ($z > 0.6$). Error bars are the standard error of the sample proportion in the binomial standard deviation. The vertical line ilustrate the $f_{red}$ inside of $R_{200}$ for the two samples.}
\label{fig6}
\end{figure} 

\begin{figure}
\includegraphics[width=86mm,height=94mm]{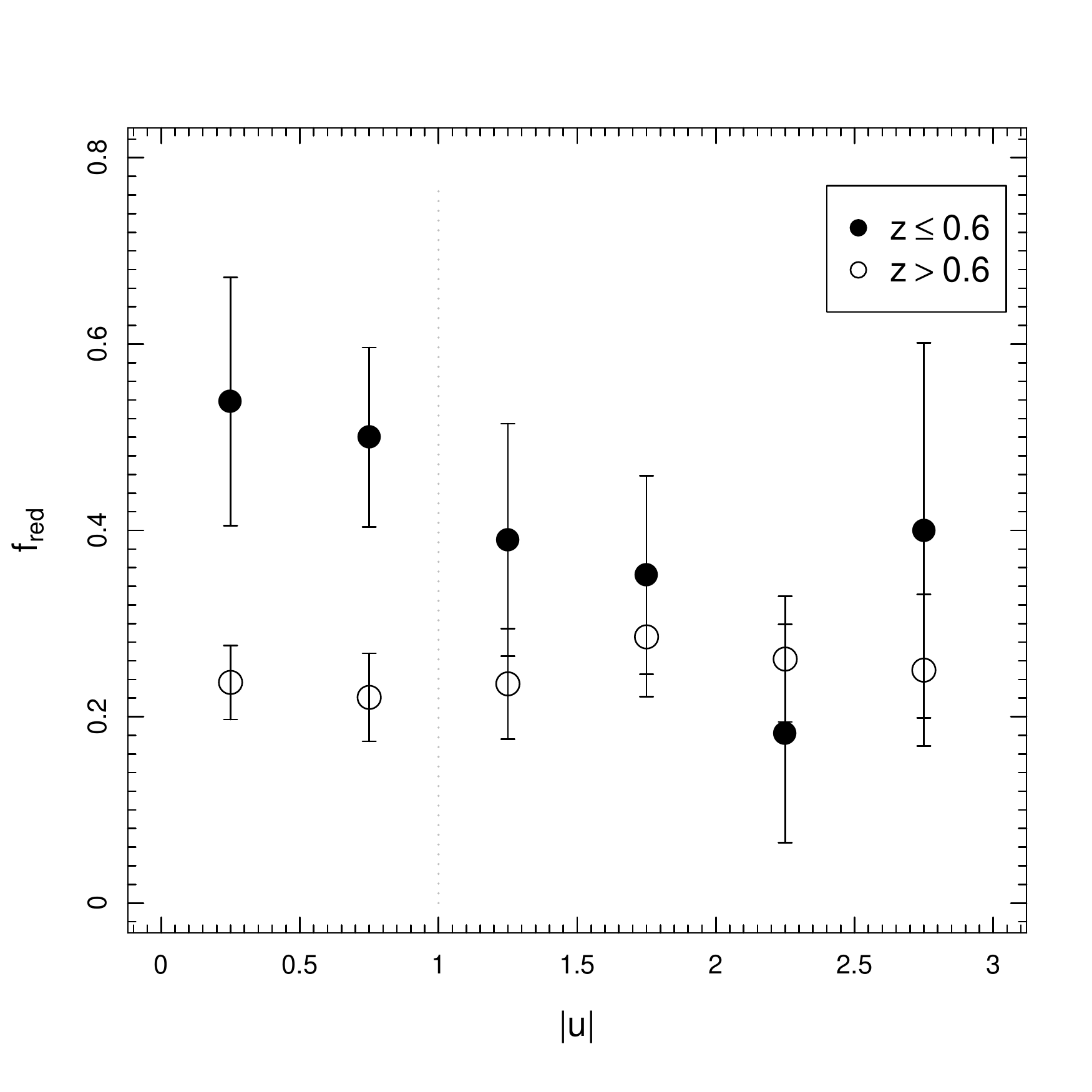}
 \caption{Fraction of red galaxies as a function of $|u|$,  taking galaxies up to $2R_{200}$ with $M_B \leq -18.5$. Filled (open) circles depict galaxies at $z\leq 0.6$ ($z > 0.6$). Error bars are the standard error of the sample proportion in the binomial standard deviation.}
\label{fig7}
\end{figure}

Using color as a proxy for galaxy type, according to the color $(U-B)_0$ separators defined in Section 3.1, we show in Fig.~\ref{fig5} the fraction of red galaxies  up to $2R_{200}$  as a function of $M_B$. Note in this figure that, for the low-{\it z} sample, the subset of galaxies brighter than $M_B\sim -19.5$ has a much higher fraction of red galaxies than the subset of fainter objects. A similar but less accentuated effect is observed for the high-{\it z} sample. At the same time, the radial variation of the red fraction indicates that there are more red objects within $R < R_{200}$ in the low-{\it z} sample than in the high-{\it z} sample, as we can see in Fig.~\ref{fig6}. In the outskirts, the  difference between the samples is much smaller. Finally, in Fig.~\ref{fig7} we see that the fraction of red galaxies is decreasing in the low-{\it z} sample up to $|u|=2.25$,
while it is nearly flat for the high-{\it z} sample. Also in this figure, note the significant higher fraction of red objects with low velocities $|u|<1$ in the low-{\it z} sample. These combined results indicate that brighter and redder objects have lower velocities and are more central than the brighter and bluer ones, thus producing the segregation effect observed in Fig.~\ref{fig4}.

\subsection{Statistical tests}

\begin{figure}
\includegraphics[width=86mm,height=94mm]{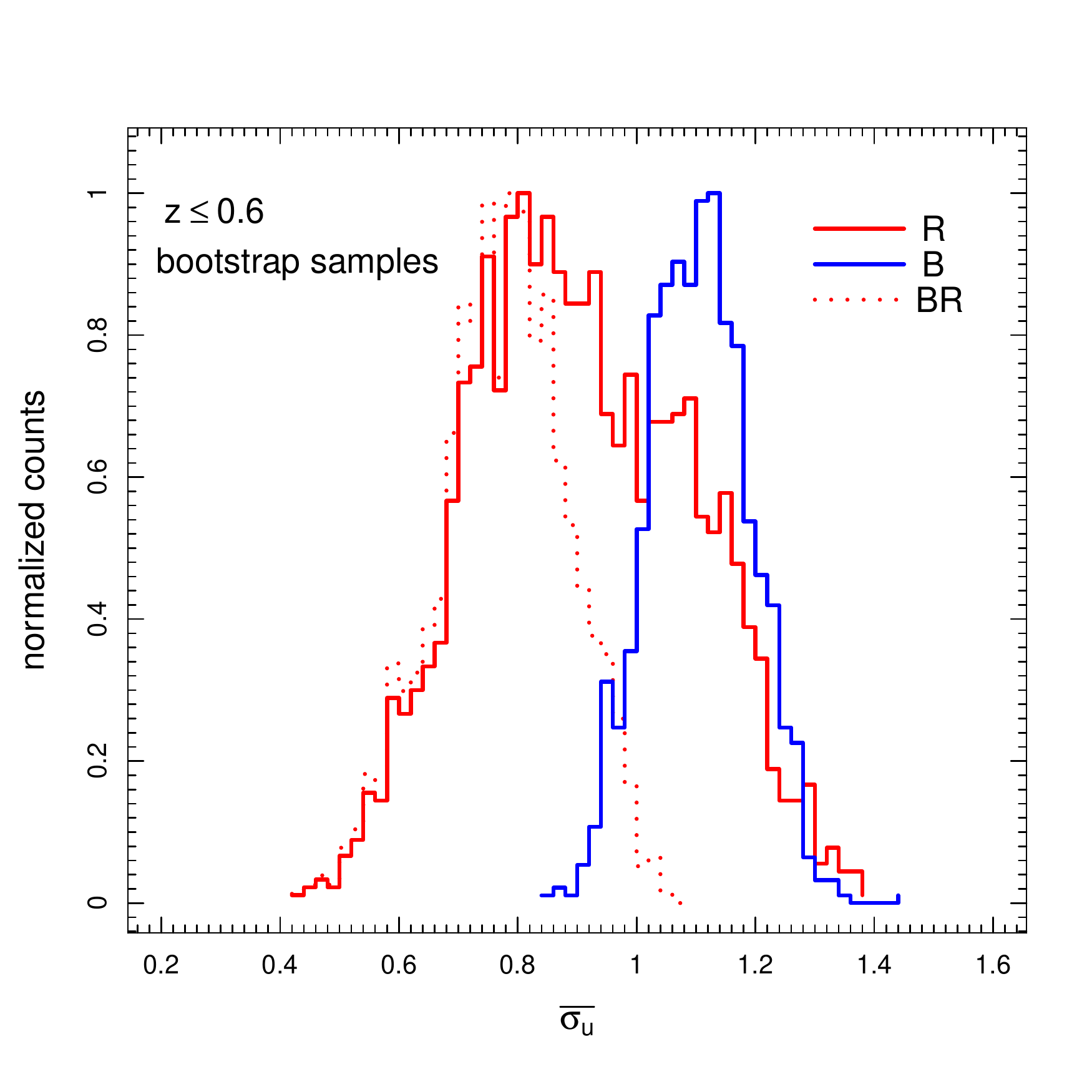}
 \caption{Distribution of $\sigma_u$ means for 1000 bootstrap samples generated for all red galaxies ($M_B < -18.5$) -- red solid lines; bright red galaxies ($M_B < -19.5$) -- red dotted lines; and all blue galaxies -- blue solid lines.}
\label{fig8}
\end{figure}

Now, let's take a closer look of the segregation effect observed for the low-{\it z} sample in comparison with the high-{\it z} sample.
Segregation has been reported to happen when redder and brighter galaxies are more clustered and lie closer to the group center both in position and in velocity than bluer and fainter galaxies \citep[e.g.][]{gi2, AG}. This is consistent with the results obtained in Section 3.2. Hence, we should pay attention to the statistical behaviour of the bright red population. To reinforce our results and verify if the distributions of $\sigma_u$ per galaxy type present significant differences with respect to the magnitudes, we run a pairwise bootstrap test for three subsamples: (i) all red galaxies (R); (ii) bright red galaxies (BR) ($M_B < -19.5$)\footnote{We define bright red objects according to Fig.~\ref{fig5}.}; and (iii) blue galaxies (B). We select 1000 bootstrap samples from our data respecting both subsample size and galaxy type (R, BR, or B). The procedure consists in comparing $k$ population means performing the hypothesis tests: $H_0: \mu_i = \mu_j , ~, i\neq j = 1,2,...,k$. For each bootstrap sample the differences $\delta_{ij}=|\bar{\sigma}_u(i) - \bar{\sigma}_u(j)|$ are computed and stored (with $\mu_i=\bar{\sigma}_u(i)$). The p-value of the test is defined as $(\sum \delta_{ij} \geq \delta_{ij}^{\rm obs})/N$, where $N$ is the number of bootstrap samples and $\delta_{ij}^{\rm obs}$ are the differences of means obtained from the original data \citep[e.g][]{wy}. By applying this procedure, we find that R galaxies have $\bar{\sigma}_u^{\rm R}=0.883\pm 0.101$ and move more slowly than B galaxies, which have $\bar{\sigma}_u^{\rm B}=1.108\pm 0.087$. The pairwise bootstrap test indicates the result is significant at the 90\% c.l. (p=0.0982). A more marked difference is found for the comparison between BR galaxies, which have $\bar{\sigma}_u^{\rm BR}=0.776\pm 0.099$, and the blue galaxies. In this case, the test indicates that BR objects are moving more slowly than B galaxies at the 99\% c.l. (p=0.0015).
In Fig.~\ref{fig8}, we present the distributions of $\sigma_u$ means for the 1000 bootstrap samples. Note in this figure the different ranges for each galaxy type: $0.43 \lesssim \bar{\sigma}_u^{\rm BR} \lesssim 1.07$, $0.43 \lesssim \bar{\sigma}_u^{\rm R} \lesssim 1.38$, and $0.84 \lesssim \bar{\sigma}_u^{\rm B} \lesssim 1.43$. While BR galaxies dominate the low velocity range, $\sigma_u \lesssim 0.8$, B galaxies have an opposite behaviour, gradually becoming dominant at $\sigma_u \gtrsim 1.0$ (together with faint red galaxies). These results reinforce the idea of  kinematic segregation  related to the galaxy type distribution in groups.

\begin{figure}
\includegraphics[width=86mm,height=94mm]{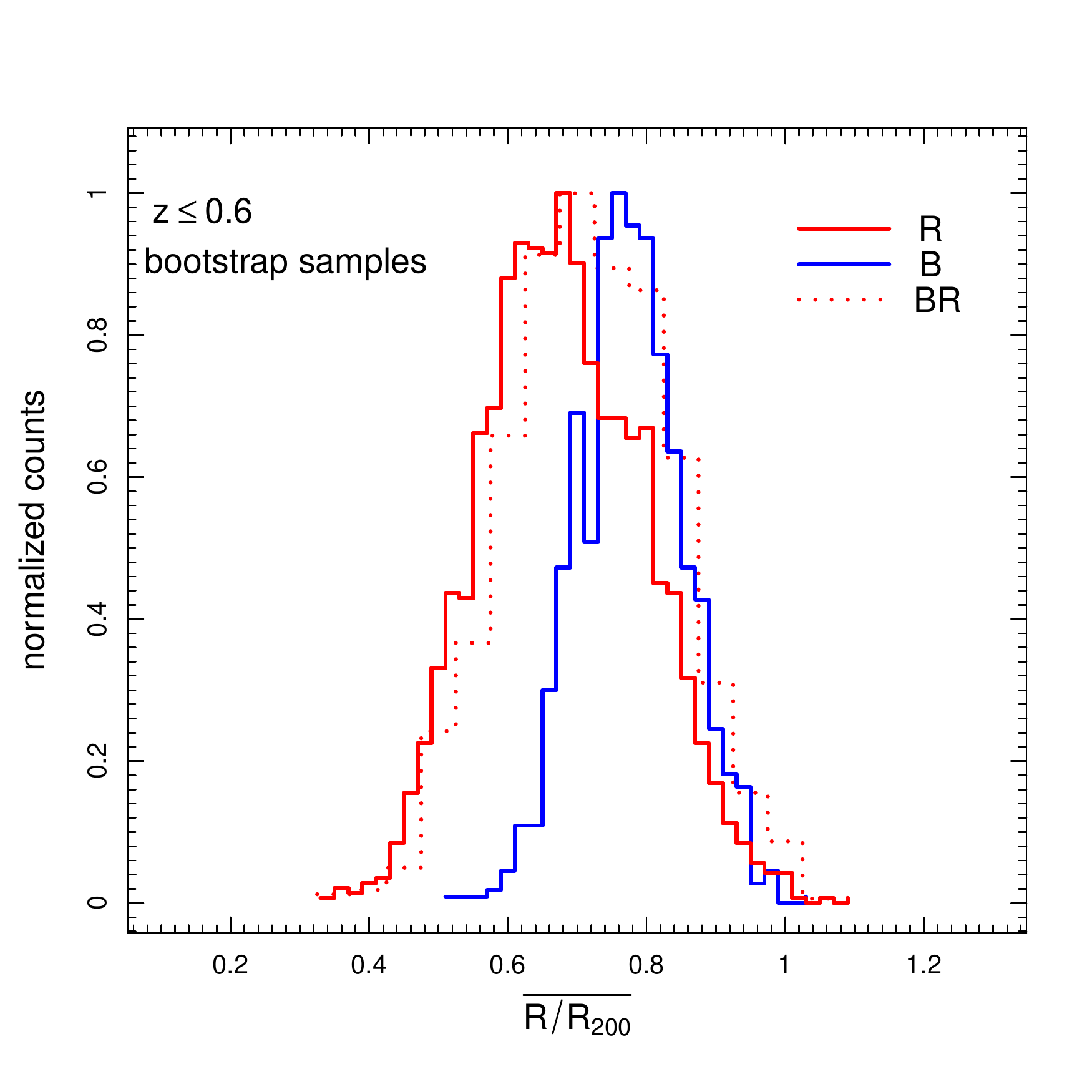}
 \caption{Distribution of ${\rm R/R_{200}}$ means for 1000 bootstrap samples generated for all red galaxies -- red solid lines; bright red galaxies ($M_B < -19.5$) -- red dotted lines; and blue galaxies -- blue solid lines.}
\label{fig9}
\end{figure}

In addition, we should consider the groupcentric distance distribution of members per galaxy type. For this aim, the boostrap test was performed again for the three subsamples, now probing the respective distributions of $R/R_{200}$ means.The bootstrap distributions are presented in Fig.~\ref{fig9}, with the following ranges: $0.33 \lesssim (\bar{R}/R_{200})^{\rm R} \lesssim 1.08$, $0.35 \lesssim (\bar{R}/R_{200})^{\rm BR} \lesssim 1.08$, and $0.52 \lesssim (\bar{R}/R_{200})^{\rm B} \lesssim 1.02$. Although a higher fraction of red galaxies is found closer to the group centers (see Fig.~\ref{fig6}), now the $\bar{R}/R_{200}$ distributions present larger overlapping areas and, unlike the previous case, the bootstrap test cannot reject the hypothesis of same means for R and B galaxies at a high confidence level. We find $(\bar{R}/R_{200})^{\rm R}=0.665\pm 0.107$, $(\bar{R}/R_{200})^{\rm BR}=0.692\pm 0.108$, and $(\bar{R}/R_{200})^{\rm B}=0.767\pm 0.095$, with the pairwise comparisons leading to  p=0.3967 (BR vs B), and p=0.2931 (R vs B). Despite the visual difference between the distributions, in the sense that red galaxies are a little more concentrated than the blue ones, it is not statistically significant for the difference of means (at best, we get $\sim$70\% c.l. comparing B and R galaxies), and thus we cannot say that red and blue galaxies are decidedly segregated with respect to  their projected groupcentric distances, although they are kinematically distinct objects at the 90\% c.l. (R vs B), or at the 99\% c.l. (BR vs B). 

\begin{figure}
\includegraphics[width=86mm,height=94mm]{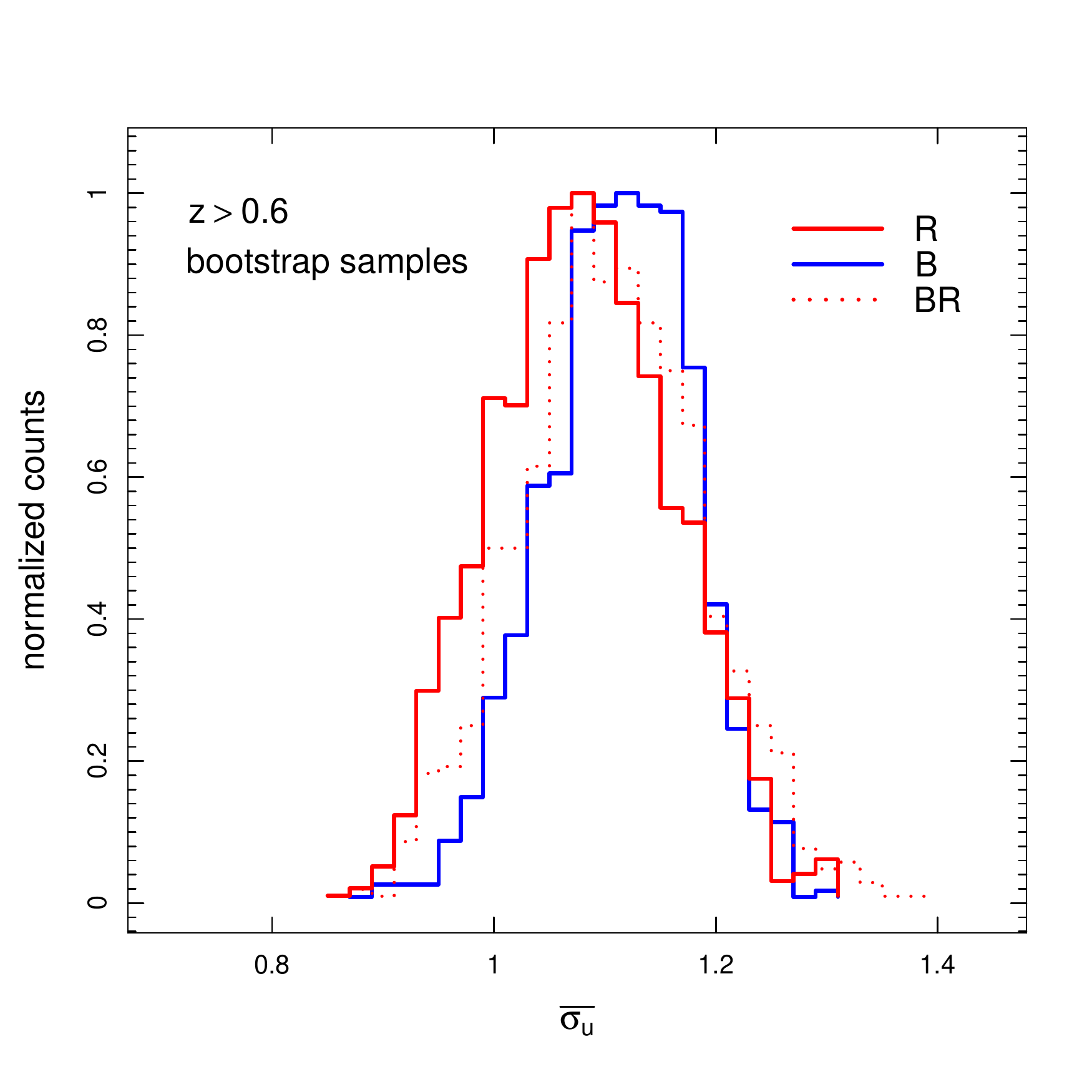}
 \caption{Distribution of $\sigma_u$ means for 1000 bootstrap samples generated for all red galaxies -- red solid lines; bright red galaxies ($M_B < -19.5$) -- red dotted lines; and blue galaxies -- blue solid lines.}
\label{fig10}
\end{figure}

\begin{figure}
\includegraphics[width=86mm,height=94mm]{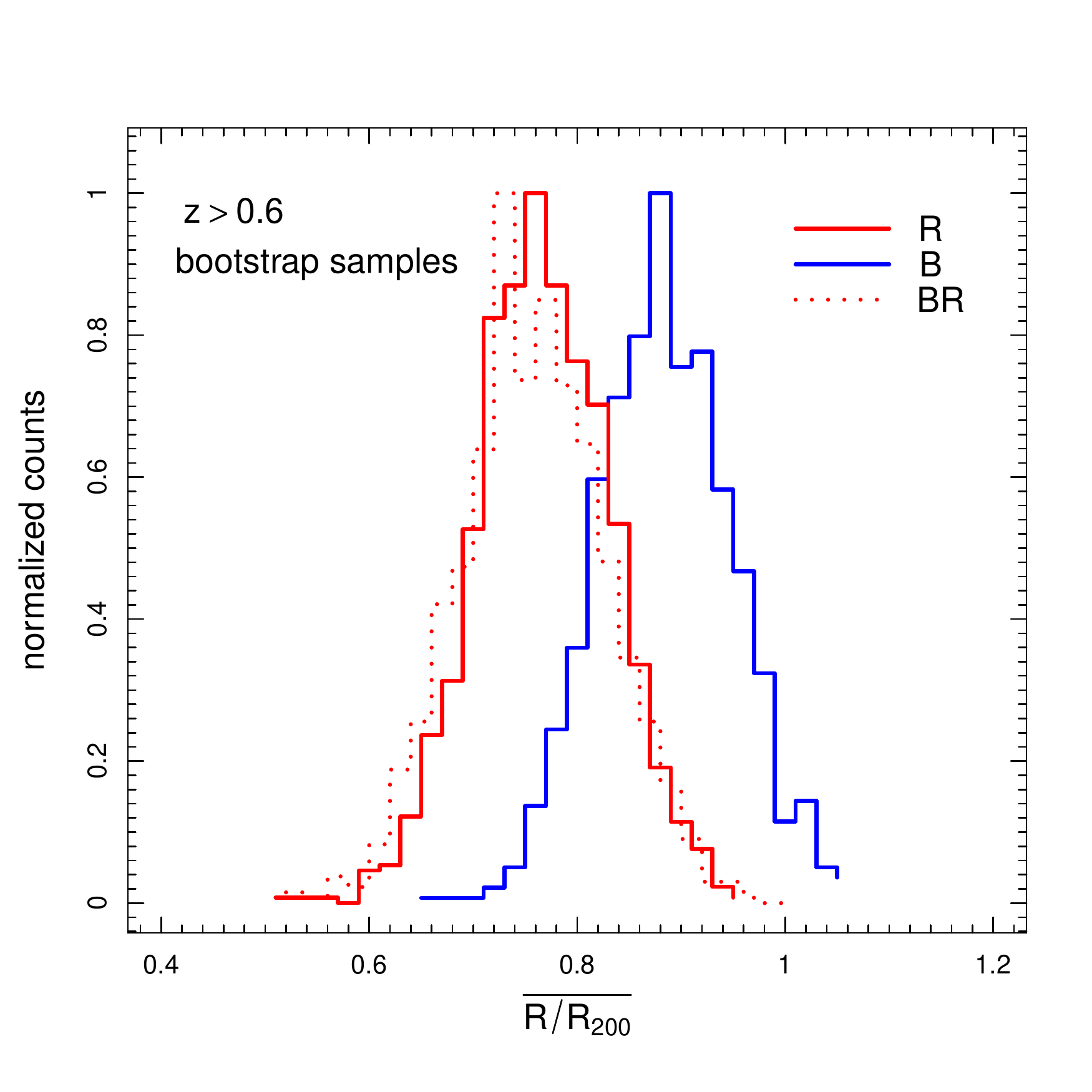}
 \caption{Distribution of ${\rm R/R_{200}}$ means for 1000 bootstrap samples generated for all red galaxies -- red solid lines; bright red galaxies ($M_B < -19.5$) -- red dotted lines; and blue galaxies -- blue solid lines.}
\label{fig11}
\end{figure}

By repeating the procedure for the high-{\it z} sample, we find that B galaxies have $\bar{\sigma}_u^{\rm B}=1.105\pm 0.066$, R galaxies have $\bar{\sigma}_u^{\rm R}=1.071\pm 0.080$, and BR galaxies have $\bar{\sigma}_u^{\rm BR}=1.090\pm 0.087$. 
The bootstrap test cannot reject the hypothesis of same $\sigma_u$ means for R and B galaxies (p=0.4389), or BR and B  (p=0.5129). Thus, there is no kinematic segregation for galaxies in the high-{\it z} sample (see Fig.~\ref{fig10}). At the same time, the bootstrap samples of $R/R_{200}$ means reveal a significant difference between red and blue galaxies, as we can see in Fig.~\ref{fig11}. We find $(\bar{R}/R_{200})^{\rm R}=0.757\pm 0.064$, $(\bar{R}/R_{200})^{\rm BR}=0.701\pm 0.088$, and $(\bar{R}/R_{200})^{\rm B}=0.898\pm 0.063$, with the pairwise comparisons leading to p=0.0211 (BR vs B), and p=0.0297 (R vs B). Hence, red and blue galaxies are  segregated in groupcentric distances at the 95\% c.l., with the red ones being more concentrated. In conjunction with the result for the low-{\it z} sample, we have a scenario where red galaxies are more central than blue galaxies at $z > 0.6$, with no kinematic segregation between them, and a different situation at $z \leq 0.6$, where there is a less compelling evidence for radial segregation between red and blue galaxies, but with the red objects (especially the bright red ones) being kinematically segregated.

\subsection{Galaxy evolution}

\begin{figure}
\includegraphics[width=86mm,height=96mm]{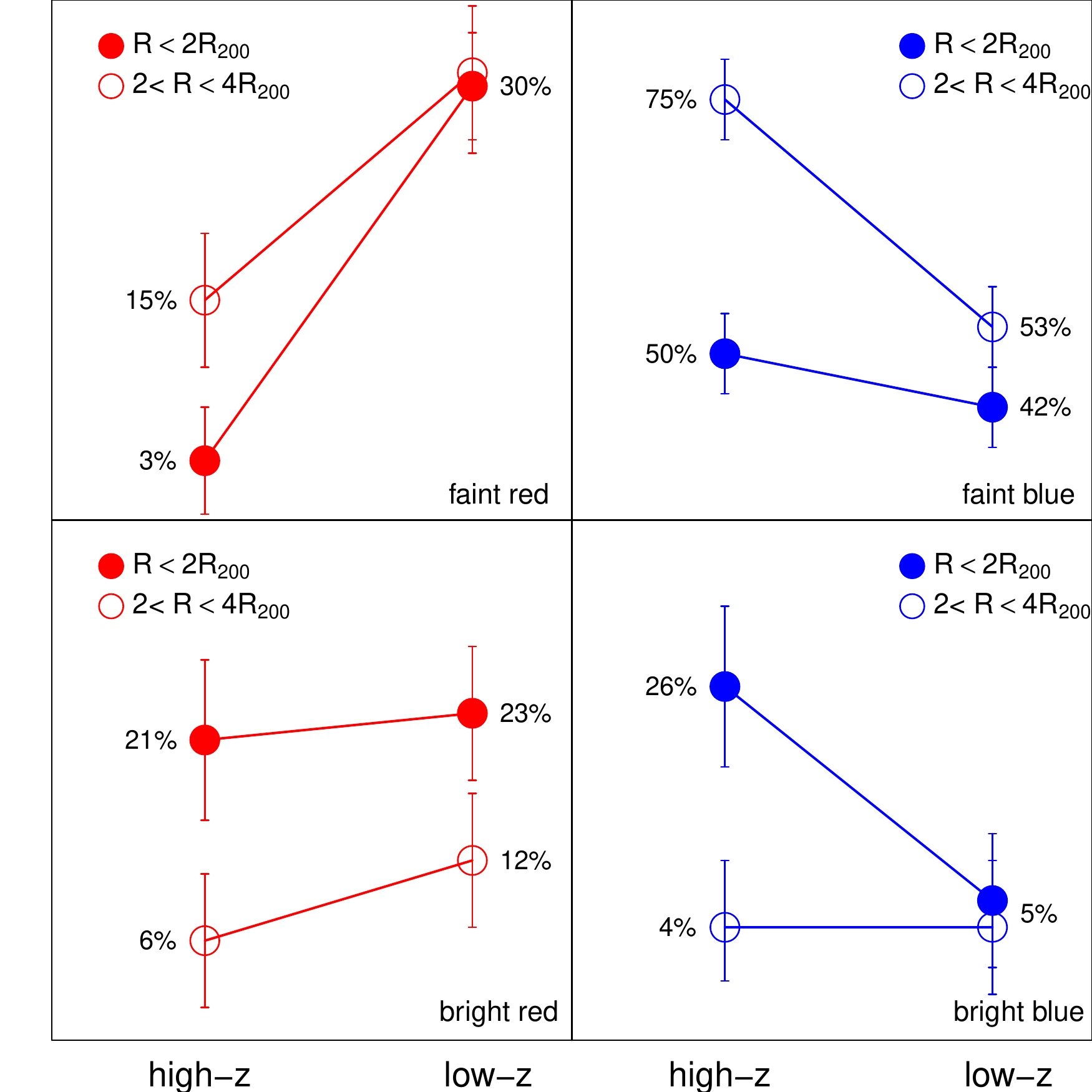}
 \caption{Galaxy evolution from high ($z\sim 0.81$) to low ($z\sim 0.39$) redshift samples, within the regions $R\leq 2\;R_{200}$ and $2 < R\leq 4\;R_{200}$.}
\label{fig12}
\end{figure}

Complementing our results, we set up a rough scheme for galaxy type evolution from high to low redshift samples, within the regions $R\leq 2\;R_{200}$ and $2 < R\leq 4\;R_{200}$ (see  Fig.~\ref{fig12}). The first point to note in this figure  is the impressive increase of faint red galaxies from high-{\it z} to low-{\it z}, in the central parts (3\% to 31\%) and also on the periphery (15\% to 32\%). This is in parallel to a more modest increase in the fraction of bright red galaxies (21\% to 23\% in the center) and (6\% to 12\% in the outskirts), and the fact that the fraction of central bright red galaxies is significantly higher in the high-{\it z} sample. At first, these findings seem to support a scenario in which low-mass galaxies are less efficient in quenching star formation in low-mass halos \citep[e.g][]{BT, cw, Poz},\footnote{Presuming that most faint objects have low masses.} in agreement with several studies reporting the existence of a deficit of galaxies at the faint end of the red sequence in high-redshift clusters \citep{DL,GB,CCS,Bil,Rud,Fass}. However, note that the total change in the faint blue sample was about $30\%$, with a small decline within $2\;R_{200}$ (50\% to 42\%), and a significant decrease for objects outside this radius (75\% to 53\%). At the same time, the total change in the faint red sample was of $45\%$, a considerably higher variation. In addition, note in Fig.~\ref{fig12} that the fraction of faint red galaxies is significantly higher in the periphery than in the central part for the high-{\it z} sample, and that this fraction becomes similar to that of the central faint red galaxies in the low-{\it z} sample. All this suggests we are not seeing just a downsizing effect, and might indicate an additional environmental quenching process driven by some mechanism favouring the appearance of faint red objects in the outskirts at $z\sim 0.8$. Indeed, by defining a field sample at this redshift (3435 galaxies with $\Delta z \geq 0.06$ and $R > 4$ Mpc away from DEEP2 groups), we find that the faint red fraction is $\sim$2\% in the field, similar to that we find in the inner parts of high-{\it z} groups. This is consistent with the scenario where the decrease in star formation and color transformation sets in at several virial radii at $z\sim 1$ \citep{Le,Gom, Bah}. Indeed, we note an important decrease of the faint blue population in the outskirts from high to low-{\it z} (75\% to 53\%). This also suggests that quenching of star formation could begin to ocurr in the infall region with part of the red fraction in the central region at lower redshifts being due to the lag between the start of quenching and the time for its effects to become apparent \citep[e.g.][]{Just}. Since galaxies move $\sim$ Mpc distances over $\sim$ Gyr timescales \citep[e.g.][]{Blg}, after $\sim$3 Gyr, galaxies initially in the outer parts of high-z sample could have migrated, becoming part of the central regions of groups at lower redshifts. As shown by \citet{Wz} half of satellites in the mass range ($M^\ast : 10^{9.7} - 10^{11.3}~{\rm M_\odot}$) first fell into groups and clusters at $z\geq 0.5$, with a broad tail out to $z\geq 1$, so they typically have experienced $\geq$ 4 Gyr evolving as a satellite. These authors show results favouring a delayed-then-rapid quenching scenario, where satellite SFRs evolve unaffected for $2-4$ Gyr after infall, and then star formation is quenched with an e-folding time of $<$ 0.8 Gyr. Due to the long time delay before quenching starts, group preprocessing should play an important role in quenching satellites \citep{Wz, l2}. \citet{Tar} put forward an alternative scenario whith an exponencial quenching timescale of 3$-$3.5 Gyr, for disc galaxies with $M^\ast \sim 10^{10} ~{\rm M_\odot}$. This scenario favours gentler quenching mechanisms such as slow ``strangulation" over more rapid ram-pressure stripping. This is also consistent with the work of \citet{Peng} arguing that in the local universe most galaxies were quenched over $\sim$4 Gyr timescales by strangulation. It is worth noting here that for a small subsample of our data \citep[103 galaxies from][]{b1b}, we have $M^\ast : 10^{8.2} - 10^{11.9}~{\rm M_\odot}$, with mean stellar mass of $\sim 10^{10}~{\rm M_\odot}$, meaning that a slow quenching scenario could also be consistent with our results. 

Still considering Fig.~\ref{fig12}, we see the increase of bright red galaxies on the periphery from the high-{\it z} to the low-{\it z} sample. This increment may happen at the expense of minor merging of faint blue/red galaxies over the $\sim$3 Gyr timescale \citep[e.g.][]{Naab,Hop}, and also by secular quenching of bright blue galaxies. The slight increase of bright red galaxies in the central parts is lower than expected and could result from the fact that our low-{\it z} sample has the brightest members  less luminous than those in the high-{\it z} sample (see Section 4). Note, however, that our central region extends up to 2$R_{200}$, and hence contains a mixture of virialized, infalling and backsplash objects \citep[e.g.][]{MA11,Jaf,Om}. Taking galaxies only within 1$R_{200}$, we find that the bright red fraction presents a larger increase, from 22\% to 28\%.

\section{Possible sample selection bias}

To assess the impact of the sample size at low and high-{\it z} on the results that we found, a two-sample test with permutation was applied. This test works as follows. First of all, the difference between the mean absolute magnitude for each sample, $\Delta M$, is calculated. Next, we combined the samples in a single dataset under the null hypothesis that there is no difference between the means that are being tested. Then, we produced two samples, randomly selected, from the combined sample and a new estimate of $\Delta M$ is made. If the sample was drawn from the same parent distribution the difference between these two estimates of $\Delta M$ should be small. The process is repeated 1000 times and we check how many times the permuted difference is equal or more extreme than the observed difference. From the application of this test we found that the magnitude distribution of the subsample of galaxies in groups at low-{\it z} used in this work is considered similar to the general low redshift sample of the DEEP2 survey at the 95\% confidence level.

Another aspect to consider is the absolute magnitude range of the low and high redshift subsamples. The sample at $z \leq 0.6$ is composed of 88 galaxies with $-20.5 \leq M_B \leq -18.5$ distributed in 25 galaxy groups while the sample at $z > 0.6$ contains 589 galaxies, with $-22.0 \leq M_B \leq -18.5$, spread in 75 groups. Since both samples have the same low luminosity end, differences at low fluxes just reflect the automated binning procedure we adopt here. But the differences at the high luminosity end are real and should be taken into account. Although we have bright galaxies in both samples, the high-{\it z} sample have slightly brighter galaxies than low-{\it z} sample. This is due to the na\-tu\-ral trend of flux limited surveys. Specifically, owing to the faint apparent magnitude range of z$\sim$1 galaxies and consequently brighter $M_B^*$ at higher redshifts (see \citet{wi} to a detailed description) the galaxies included by the DEEP2 survey tend to be brighter at higher {\it z}. Besides that, as can be seen on Fig. \ref{fig2} (and exhaustively discussed on \citet{wi} and \citet{g2}) red sequence galaxies in DEEP2 will have a brighter absolute magnitude limit than blue galaxies at the same redshift. However, a larger number of bright (and massive) galaxies in the sample would hardly be responsible (alone) for wiping out the segregation effect in the high redshift subsample. 

\section{Discussion}

In this work, we searched for segregation phenomena in galaxy groups in the range of $0.2<z<1$, using a sample of groups selected from the 4th Data Release of the DEEP2  galaxy redshift survey.   The sample was divided into two stacked systems: low($z\leq 0.6$) and high ($z>0.6$) redshift groups, with members being classified in red and blue objects. Assuming that the color ${\rm U-B}$ can be used as a useful proxy for the galaxy type, we found that the fraction of blue objects is higher in the high-{\it z} sample, with blue objects being dominant at $M_{B} > -19.5$ for both samples, and red objects being dominant at $M_{B} < -19.5$ only for the low-{\it z} sample. Also, the radial variation of the red fraction indicates that there are more red objects with $R < R_{200}$ in the low-{\it z} sample than in the high-{\it z} sample. Our analysis also indicates statistical evidence of kinematic segregation for the low-{\it z} sample: redder and brighter galaxies present lower mean velocity dispersions than bluer and fainter ones. Red and blue objects, however, present less separated mean groupcentric distance distributions, with the pairwise test indicating that the red population is more concentrade only at the 70\% c.l. Interestingly, the analysis of the high-{\it z} sample reveals an opposite result: while red and blue galaxies have velocity dispersion distributions not statistically distinct, redder objects are significantly more concentrated than the bluer ones at the 95\% c.l. From the mean difference in redshifts of the two samples, we estimate that the minimum timescale for the appearance of these inverted segregation effects is approximately 3.0$\pm$0.3 Gyr. The challenge, then, is envisioning how these results can emerge in the context of galaxy evolution.

Our main result is presented in  Fig.~\ref{fig4}. To understand the difference at the bright end observed in this figure we should consider that in the low-{\it z} sample the first bins are dominated by red objects, while blue galaxies dominate all the magnitude range in the high-{\it z} sample (see Fig.~\ref{fig5}). The red galaxies in the low-{\it z} sample show lower velocity dispersions (see Fig.~\ref{fig6}), in agreement with the works of e.g. \citet{a2}, \citet{go1} and \citet{AG} for different samples of low redshifts cluster galaxies ($z< 0.1$). This suggests that brighter and redder objects are former inhabitants of the system, having experienced more environmental effects along time, and that have achieved the energy-equipartition status through dynamical interactions on a timescale of $\sim$3 Gyr since $z\sim 0.8$. The lower fraction of red/bright/low velocity objects in the high-{\it z} sample explains the difference observed at the bright end in  Fig.~\ref{fig4}. Still looking at this figure,  we see a velocity upturn only observed in the last bins of the low-{\it z} sample. This effect may indicate a fraction of faint blue galaxies which entered into $2R/R_{200}$ before a significant quenching has happened. Their higher velocity offsets would have acquired as they approach the group core, falling into the gravitational potential \citep[e.g.][]{Fal,Jaf}. The absence of similar objects in the high-{\it z} sample suggests that infalling faint blue galaxies have not yet travelled a long journey across the group at $z\sim 0.8$.

Other important result we reported in Section 3.3 is the reversing behaviour of red and blue galaxies with respect to velocity  and groupcentric distances segregation, with redshift. Regarding velocity segregation, the preceding paragraph provides a qualitative scenario. Now, to explain the spatial segregation, we shoud notice that our analyses in Sections 3.2 and 3.3 take account galaxies within $2R/R_{200}$. One can reasonably assume that such objects at lower redshifts correspond to a mixture of descendants of galaxies at higher redshifts in the same radii and of infalling objects from outer radii. Thus, both survival and replenishment of galaxies should be expected over the time, and two important factors come into play: (i) the accretion rate of galaxies; and (ii) the orbital dependence of galaxy properties \citep[e.g.][]{Bi4,ID}. Indeed, regarding velocity segregation, it has also been interpreted as red and blue galaxies having different kinds of orbits, with the orbits of blue galaxies being more anisotropic than the red ones \citep[e.g.][]{Bi4}. Recently, \citet{Bi5} verified that the anisotropy profile of $z\sim 1$ clusters is nearly isotropic near the cluster center, and increasingly elongated with radius. This result is consistent with a halo evolution through an initial phase of fast collapse and a subsequent slow phase of inside-out growth by accrection of field material \citep[e.g.][]{Lapi}. Since the accretion rate of galaxies from the field is higher at higher redshifts \citep[e.g.][]{Mc09}, our sample at $z\sim 0.8$ is expected to be more affected  by recent infalls, which had less time to go deeper into the group potential. This could explain the development of a more marked difference between the mean groupcentric distance of red and blue galaxies (see Fig.~\ref{fig12}). After $\sim$3 Gyr, part of these infalling galaxies may reach the $R < 2R_{200}$ region, at $z\sim 0.4$, mixing with virialized and backsplash objects, and thus presenting a less pronounced radial segregation between red and blue galaxies. 

\section{Summary}

Our main achievements in this work are:

\begin{enumerate}
\item Velocity segregation in the low-{\it z} sample. We found a well pronounced relation between the normalized velocity dispersion, $\sigma_u$, and the absolute magnitude $M_B$, where the brighter objects are moving more slowly than the less luminous ones. Statistical tests reinforce our finding of lower velocity dispersions for redder and brighter galaxies at low redshifts at the 99\% c.l. This result is related to the higher fraction of redder/brighter/lower velocity objects in the low-{\it z} sample in comparison to the objects in the high-{\it z} sample, where no velocity segregation was verified. \item Strong spatial segregation in the high-{\it z} sample, with red galaxies being more central. Statistical tests indicate that red and blue galaxies are segregated with respect to the groupcentric distances at the 95\% c.l. For galaxies in the low-{\it z} sample there is a weaker evidence for spatial segregation between red and blue galaxies, only at the 70\% c.l. This result is probably related to the different accretion rate of galaxies in groups at different redshfits, and the time needed to galaxies go deeper into the group potential and mix with other galaxy types.
\item The timescale for velocity segregation emergence  is $\sim$3 Gyr starting (at least) from $z\sim 0.8$. This seems to be the same timescale for significant infalling of objects from the outer radii in the high-{\it z} sample to the inner radii in the
low-{\it z} sample.
\item Galaxy evolution in this same timescale is consistent with a slow star formation quenching scenario.
Our results are consistent with both pre-processing and slow strangulation processes. 
\end{enumerate}

Future work will include complete stellar mass information, and then mass segregation and other aspects of the problem will be deeply investigated in the two samples.

\section*{Acknowledgements}

RSN thanks the support from CAPES PhD grant. PAAL thanks the support of CNPq, grant 308969/2014-6.
ALBR would like to thank the financial support from the project Casadinho PROCAD - CNPq/CAPES number 552236/2011-0. ALBR also thanks the support of CNPq, grant 309255/2013-9.






\label{lastpage}
\end{document}